\documentclass{article}
\usepackage{graphicx}
\usepackage{epsfig}
\usepackage{amsfonts}
 \usepackage{amssymb}
\usepackage{enumitem}
\usepackage{cite}

\usepackage{bm,color}

\date{\today}

\newcommand{\insertplot}[5]{\begin{figure}
 \hfill\hbox to 0.05in{\vbox to #5in{\vfill
 \inputplot{#1}{#4}{#5}}\hfill}
 \hfill\vspace{-.1in}
 \caption{#2}\label{#3}
 \end{figure}}
 \newcommand{\inputplot}[3]{
 \special{ps: plotfile #1}
}
\newcounter{fig}

\def\eqref#1{Eq.~(\ref{#1})}

\def\Eq#1{\begin{equation} #1 \end{equation}}
\def\Eqr#1{\begin{eqnarray} #1 \end{eqnarray}}

\newcommand{\ee}{\end{equation}}
\newcommand{\eea}{\end{eqnarray}}
\newcommand{\be}{\begin{equation}}
\newcommand{\bea}{\begin{eqnarray}}

\newcommand{\e}{\mbox{e}}

\newcommand{\nn}{\nonumber}

\def\X5sp{{\rm X}_5}
\def\Y3sp{{\rm Y}_3}
\def\Z3sp{{\rm Z}_3}

\def\lap{{\triangle}}
\def\e{{\rm e}}

\begin{document}

\title{\bf De-singularizing the extremal GMGHS black hole 
\\
via higher derivatives corrections } 

\author{
{\large Carlos Herdeiro}$^{\dagger}$,
{\large Eugen Radu}$^{\dagger}$ 
{\large}and {\large Kunihito Uzawa}$^{\ddagger}$  
\\ 
\\
$^{\dagger}$ {\small Departamento de Matem\'atica da Universidade de Aveiro and } 
\\
 {\small  Centre for Research and Development  in Mathematics and Applications (CIDMA),}   
\\
   {\small Campus de Santiago, 3810-183 Aveiro, Portugal}
	\\
	$^{\ddagger}${\small Department of Physics,
School of Science and Technology,} 
\\
 {\small 
Kwansei Gakuin University, Sanda, Hyogo 669-1337, Japan} 
}

\date{February 2021}

\maketitle

 \begin{abstract} 
The Gibbons-Maeda-Garfinkle-Horowitz-Strominger (GMGHS) 
black hole is an influential solution of the low  energy 
heterotic string theory. As it is well known, 
it presents a  singular extremal limit.  {We construct a 
regular extension of the GMGHS extremal black hole 
in a model
with $\mathcal{O}(\alpha')$ corrections in the action, by solving the fully non-linear equations of motion.
The 
de-singularization is supported by the $\mathcal{O}(\alpha')$-terms.
The regularised extremal GMGHS  BHs are asymptotically flat, 
possess a  regular (non-zero size) horizon of spherical topology,
with  an
$AdS_2\times S^2$ near horizon geometry, and their entropy is 
proportional  to the electric charge. The near horizon solution is obtained analytically  and some illustrative bulk solutions are constructed numerically.} 
 \end{abstract}


\section{Introduction}

Low energy string theory compactified to four spacetime dimensions admits a famous black hole (BH) solution, found by Gibbons and Maeda~\cite{Gibbons:1987ps} and, independently,
 by Garfinkle, Horowitz and Strominger~\cite{Garfinkle:1990qj} --  
 from now on dubbed the GMGHS BH.
This solution can be described either in
the Einstein frame or  in the conformally related
string frame. 
In the former case (which is the one considered in this work), 
the model's action is the sum of an Einstein term,
a kinetic term for a scalar field (the dilaton)
and a Maxwell term with an exponential coupling to the dilaton  - see~\eqref{Ef} below.
The GMGHS BH is the simplest solution of this model,
representing a charged, static and spherically symmetric horizon surrounded by scalar hair.
This stringy extension of the Reissner-Nordstr\"om (RN) BH has
attracted an enormous attention, finding
 a variety of interesting applications, $e.g.$~\cite{Garfinkle:1990qj, Gross:1986mw, Wiltshire:1988uq, 
Horne:1992bi, Gregory:1992kr, 
Giddings:1993wn, Dowker:1993bt, Natsuume:1994hd, Campanelli:1994sj,Horowitz:1997uc,Matos:2000za,Mars:2001pz,Chen:2004zr,
Kats:2006xp, Li:2014fna, Zhang:2015jda, Li:2015bfa, Delgado:2016zxv, Cano:2018qev, 
Cheung:2018cwt, Ovgun:2018prw, Herdeiro:2019yjy, Cheung:2019cwi, Astefanesei:2019pfq, Loges:2019jzs, Cano:2019oma, 
Cano:2019ycn,Blazquez-Salcedo:2019nwd,Astefanesei:2019qsg}.

An awkward property of the GMGHS solution is that
 its extremal limit is  singular:
 when taking this limit the area of the spatial sections of the horizon shrinks to zero and the Kretschmann scalar blows up at the (would be) horizon.
Since neither Ref. \cite{Gibbons:1987ps} nor \cite{Garfinkle:1990qj} 
 take into account the possible stringy $\alpha'$-corrections to the Einstein-Maxwell-dilaton action one may ask  if such corrections could  de-singularize the extremal solution.

A {\it perturbative} extension of the extremal {\it magnetic} GMGHS 
 BH has been constructed by Natsuume in Ref. \cite{Natsuume:1994hd}  (to first order
order in $\alpha'$).  
As found therein \cite{Kats:2006xp, Cheung:2018cwt, Cano:2018qev, 
Cano:2019oma, Loges:2019jzs, Cano:2019ycn},
 the corrected solution inherits all basic properties of the extremal GMGHS BH;
in particular, the horizon area still vanishes.
{On the other hand, to the best of our knowledge,  the task of constructing the \textit{fully non-linear} BH solutions of the $\mathcal{O}(\alpha')$ corrected  action has not yet been considered in the literature. This is   presumably due to the complexity of the field equations. Yet, such construction can reply to the key question whether such corrections can desingularise the extremal GMGHS solution.}
The main purpose of this work is to report results in this direction. 
Starting with a general model for the  $\mathcal{O}(\alpha')$ corrections to the Einstein-Maxwell-dilaton action
(which is essentially the one in Ref. \cite{Natsuume:1994hd}),
we find  that 
the field equations of the full model
possess an exact solution describing a Robinson-Bertotti-type vacuum,
with an $AdS_{2}\times S^2$ metric, an electric field and a constant dilaton.
On the other hand, we also find there is  no counterpart of this solution with a 
magnetic charge.

The  supergravity action includes, in general, a 
tower of corrections of all powers in $\alpha'$ because of the 
recursive definition of the Kalb-Ramond 3-form field strength, 
which breaks the supersymmetry in the supergravity theory. 
The term of quadratic order in curvature is obtained imposing 
supersymmetry of the theory at first order in $\alpha'$ if we 
consider the Chern-Simons term in the field strength. 
There are further  higher power corrections in the curvature
of the torsionful spin connection, which are required to preserve 
supersymmetry order by order \cite{Bergshoeff:1989de}. 
Additional higher-curvature 
corrections unrelated to the supersymmetrization of the Kalb-
Ramond kinetic term are also present, although those appear 
first at cubic order in $\alpha'$. 
These additional higher-order (like ${\alpha'}^2$, ${\alpha'}^3$ 
$\cdots$) corrections may drastically modify the non-perturbative 
result obtained from our setup. 
The properties of the solutions with higher curvature corrections, within the heterotic 
string theory, is still largely an uncharted territory. 
Although the conditions imposed by supersymmetry have been studied 
in detail, the solution of the Einstein equations are not well known yet, 
especially for the interesting case of the heterotic string with 
fluxes. 

As for the extremal RN  solution in the Einstein-Maxwell model,  we expect this ``attractor" to describe the near-horizon geometry of an extremal BH with a regular horizon.
That is, the $\mathcal{O}(\alpha')$-corrections de-singularize the extremal limit 
of the GMGHS BH, leading to a non-zero size, regular horizon\footnote{ 
Finding the local solutions in the vicinity of the (extremal) horizon
	($i.e.$ the attractor)
	does not guarantee the
existence of global  solutions with the right asymptotics. 
Progress in this direction  
 requires an explicit construction of the bulk extremal BHs.}.
It is well known that such near horizon geometry is  a key  feature of static supersymmetric BHs, providing the attractor
mechanism by which horizon scalar fields values  are determined by the charges carried by the BH and insensitive to the asymptotic values of the
scalar fields~\cite{Sen:2004dp, Sen:2005kj, Sen:2005wa, Sen:2005iz}.
The entropy of these BHs is consistent with the  microscopic states counting of the associated $D$-brane system.
These are described by the inclusion of higher derivative corrections 
in the generalized prepotential together with the  supergravity low energy  description~\cite{Maldacena:1997de, 
deWit:1996gjy, Behrndt:1998eq}.

In this work
we shall provide numerical evidence for the existence of  non-perturbative extensions of the 
GMGHS BHs which are extremal, asymptotically flat and regular, on and outside the horizon.
These solutions represent global (bulk) extensions of the aforementioned attractors found analytically and possess a variety of interesting
properties. 
For example, 
as for the attractors,
the entropy of the extremal BHs is proportional to the electric charge. 
Also, their charge to mass ratio is always greater than one.

This paper is organised as follows. 
In Section 2 we describe the full model,  including  $\alpha'$ 
corrections and describe briefly the GMGHS  solution. 
In  Section  3 we  consider the $\alpha'$ corrected solutions; 
first we construct (analytically) the  attractors and then we give (numerically) 
 examples  of global bulk solutions.
 Section 4 gives some concluding remarks. 
In the appendix, we summarize some formulas, 
including the equations of motion and our static spherically symmetric 
ansatz, which are needed for the results in section 3.

\section{The model }

\subsection{The action}

The starting point is the string-frame action including  
order $\alpha'$ terms
\begin{eqnarray}
\label{sf}
S_s =\int d^4x\sqrt{-\tilde{g}}\e^{-2\phi}
\bigg[
 \tilde{R} 
+4\left(\tilde{\nabla}\phi\right)^2- F^2
+\alpha' 
\bigg  \{
a\left(\tilde{R}_{\mu\nu\rho\sigma}
\tilde{R}^{\mu\nu\rho\sigma}-4\tilde{R}_{\mu\nu}\tilde{R}^{\mu\nu}
+\tilde{R}^2\right)
\\ 
\nonumber
+b\left(F^2\right)^2
+c F^2\left(\tilde{\nabla}\phi\right)^2
+h\tilde{R}^{\mu\nu\rho\sigma}F_{\mu\nu}F_{\rho\sigma}
\bigg\}
\bigg],
\end{eqnarray}
where $\tilde{g}$\,, $\tilde{R}$\,, $\tilde{R}_{\mu\nu}$\,, 
$\tilde{R}_{\mu\nu\rho\sigma}$\,, $\tilde{\nabla}$ 
denote determinant, Ricci scalar, Ricci tensor, 
Riemann curvature tensor, covariant derivative constructed 
from the string frame metric $\tilde{g}_{\mu\nu}$\,, 
respectively. 
Also, $F=dA$ is the $U(1)$ field strength tensor
and $\phi$ is a real scalar field: the dilaton. $a,b,c,h$ are constant coefficients.

The corresponding expression of the action in the  Einstein frame 
(which is the case of interest in this work)
is found via a conformal transformation,  with
\Eq{
\tilde{g}_{\mu\nu}=\e^{2\phi}g_{\mu\nu}\,. 
  \label{EA:ct:Eq}
}
This results in  ($g_{\mu\nu}$ being  the Einstein frame metric):
\begin{eqnarray}
\label{Ef}
S_E=\int d^4x\sqrt{-g}
{\cal L}, \qquad 
~~{\rm where }~~
{\cal L}=
 R
-2\left(\nabla\phi\right)^2- \e^{-\phi}F^2
+\alpha'{\cal L}'~,
\end{eqnarray}
with the leading order corrections in $\alpha'$
\Eqr{
{\cal L}'&=&
a~e^{-\phi}\left[R_{\mu\nu\rho\sigma}
R^{\mu\nu\rho\sigma}-4R_{\mu\nu}R^{\mu\nu}
+R^2-2R\lap\phi+4\nabla_\mu\nabla_\nu\phi
\nabla^\mu\nabla^\nu\phi-\left(\lap\phi\right)^2
\right.\nn\\
&&\left.
+3\lap\phi\left(\nabla\phi\right)^2
+\frac{3}{4}\left(\nabla\phi\right)^4
+2R_{\mu\nu}\left(2\nabla^\mu\nabla^\nu\phi
-\nabla^\mu\phi\nabla^\nu\phi\right)\right]
\nn\\
&&
+b~\e^{-3\phi}\left(F^2\right)^2 
+\frac{c}{4}\,\e^{-2\phi}F^2\left(\nabla\phi\right)^2
+h~e^{-2\phi}F_{\mu\nu}F_{\rho\sigma}\left[R^{\mu\nu\rho\sigma}
+g^{\sigma\left[\mu\right.}\nabla^{\left.\nu\right]}\nabla^\rho\phi
-g^{\rho\left[\mu\right.}\nabla^{\left.\nu\right]}\nabla^\sigma\phi
\right.\nn\\
&&\left.+\frac{1}{2}\left\{\left(\nabla^{\left[\mu\right.}\phi\right) 
g^{\left|\sigma\right|\left.\nu\right]}\nabla^\rho\phi
-\left(\nabla^{\left[\mu\right.}\phi\right) 
g^{\left|\rho\right|\left.\nu\right]}\nabla^\sigma\phi
-g^{\rho\left[\mu\right.}g^{\left|\sigma\right|\left.\nu\right]}
g^{\rho'\sigma'}\nabla_{\rho'}\phi\nabla_{\sigma'}\phi
\right\}
\right].
  \label{EA:ct2:Eq}
}
The variation of (\ref{Ef}) with respect to $g_{\mu\nu}$, 
$\phi$
and $A_\mu$
leads to 
the equations of motion of the model.
However, their expression is  too complicated to include here.

 Let us briefly comment  on the string theory origin of the above action.
The $U(1)$ gauge field arises as a subgroup of the 
${\rm E}_8\times {\rm E}_8$ or $SO(32)$ gauge group in the 
low-energy effective theory of the heterotic string 
\cite{Gross:1986mw}. 
We set the remaining gauge fields and the 
antisymmetric field strength $H_{\mu\nu\rho}$ to zero. 
 Then the four-dimensional action emerges from heterotic string theory compactified 
on a six-dimensional torus. 
 This includes 
corrections 
due to the next order terms such as $R^2$\,, $F^4$\,, 
$F^2\left(\nabla\phi\right)^2$ 
\cite{Natsuume:1994hd, Kats:2006xp}\,. 
After eliminating terms in the effective action 
by field redefinitions, the corrections to the 
leading order in $\alpha'$ can be expressed as 
(\ref{EA:ct2:Eq}).

The higher-order ($\alpha'$) terms in the four-dimensional 
effective action can be considered corrections due to quantum gravity~\cite{Kats:2006xp, 
Cheung:2018cwt, Deser:1974cz}. 
Let us estimate the mass scaling  of these terms, for extremal BHs, near the horizon 
$r\sim M$. 
The 
GMGHS BH solution implies
that any derivative contributes a factor of order $M^{-1}$; then,  
the curvature tensors are $R\sim M^{-2}$\,, $R^2\sim M^{-4}$\, 
and the gauge field strength gives $F
\sim M^{-1}$\,. 
Hence we have, $e.g.$ $(F^2)^2\sim M^{-4}$\,,  $F^2(\nabla\phi)^2\sim M^{-4}$\,,
$F^2\nabla^2\phi\sim M^{-4}$\, 
and $F^2R\sim M^{-4}$. 
We assume the classical mass-charge 
relation for extremal BHs  approximately holds and they are sufficiently 
 large,
$Q\simeq M\gg 1$, in units of the Planck mass 
$M_{\rm Pl}$\,. 
We  conclude that the $\alpha'$ corrections 
are suppressed by powers of $(\alpha'/M)^{2}$\,. 
Thus,  we consider 
the leading-order $\alpha'$ corrections throughout only, assuming the BHs are sufficiently macroscopic $\alpha'/M\ll  1$.

 Moreover, 
the leading order terms in the Lagrangian ${\cal L}$ have $M^{-2}$ while 
the leading order correction in the Einstein frame 
are given by (\ref{EA:ct2:Eq})
\cite{Natsuume:1994hd, Kats:2006xp};
 all terms involving $\nabla_\rho F_{\mu\nu}$ are dropped, 
without loss of generality, 
since they are equivalent, via the Bianchi identities, 
to terms already accounted for or terms involving $\nabla_\mu 
F^{\mu\nu}$~\cite{Cheung:2018cwt, Deser:1974cz}\,. 
Such terms vanish in the absence of charged matter sources, 
which is the case considered in this paper. 

We do not consider terms that are proportional to the
Kalb-Ramond three-form field strength $H$,
\Eq{
H=dB+\frac{1}{4}\left(\omega_{\rm L}
-\omega_{\rm Y}\right)\,, 
}
where $B$ is 2-form gauge potential,  
$\omega_{\rm L}$, $\omega_{\rm Y}$ denote the Lorentz and 
Yang-Mills Chern-Simons forms, 
respectively. 
The presence of $H$ in the heterotic string is required 
due to $\alpha'$ corrections in the Bianchi 
identity which are needed for anomaly cancellation.
Since 
the Chern-Simons forms act as sources for $H$\,, 
we are in general not able to simply set $H=0$\,. 
However, the Lorentz Chern-Simons form can be written by  
the exterior derivative of a three form for the spherically 
symmetric metrics. Hence, we can absorb it into the 
definition of 2-form gauge potential $B$
\cite{Campbell:1990fu}.
Moreover, the Yang-Mills Chern-Simons form vanishes for the
purely magnetic case \cite{Natsuume:1994hd}. 

\medskip

In the expression  (\ref{Ef}), $a,b,c $ and $h$
are constants which are not fixed \textit{a priori}.
The coefficient $a$ is arbitrary (and nonzero), since it can be changed by 
scaling $\alpha'$.
The choice in Ref. \cite{Natsuume:1994hd} 
(employed also here)
is
$a=1/8$.
That reference gives also an
 argument that $h=0$.

As for $b$ and $c$, to the best of our knowledge, the only concrete computation
is in 
Ref. \cite{Natsuume:1994hd}. 
The idea there was to construct the (leading order) $\alpha'$-correction 
to the {\it magnetic} GMGHS BH, in the extremal limit.
The corrected solution is supposed to possess the same near-throat behaviour and the same
far field asymptotics as the original solution    (with $\alpha'=0$).
This argument
implies
\begin{eqnarray} 
c=2 \ ,
\end{eqnarray}
(for the conventions here),
while the value of $b\geqslant 0$
is arbitrary\footnote{
Since the four-derivative terms can be  
fixed by supersymmetry, we can obtain the value of $b$  
\cite{Bergshoeff:1989de}. 
In this letter, however, we do not focus on the supersymmetric solutions.
}.

Generalizing the discussion, 
 $\alpha'-$ (or   quantum gravity) corrected Einstein-Maxwell-dilaton theories 
have attracted much attention  
\cite{Kats:2006xp, Loges:2019jzs, Cano:2019oma, 
Cano:2019ycn} 
{\it even  beyond}
 the stringy models. 
Within string theory and its low-energy 
effective field theories, the coefficients of each terms in 
the string effective action are 
invariant under field redefinitions \cite{
Natsuume:1994hd, Kats:2006xp}, which were then taken from the 
heterotic string calculations \cite{Gross:1986mw}: 
$a=1/8$ and $h=0$\,. 
The parameter $c$ can be fixed by a requirement of 
consistency with exact results that were obtained 
for the  GMGHS   BH \cite{Giddings:1993wn} while the value of
$b$ does not affect the correction to the BH mass. 
On the other hand, there are studies of BHs based on the most general collection of 
four-derivative terms for 4-dimensional Einstein-Maxwell-dilaton 
theories \cite{Loges:2019jzs}. We will here construct the $\alpha'$ corrected GMGHS 
solution within the framework of the low-energy effective theory of the heterotic string.
But our choice here is to work with some slight more generality.  Thus we take:
\begin{eqnarray}
a=\frac{1}{8}\ , \qquad b,c~~{\rm arbitrary}\ , \qquad h=0 \ .
\end{eqnarray}

\subsection{Ansatz and entropy}

In this work we are interested in static spherically symmetric solutions
with a purely electric $U(1)$ potential.
An Ansatz suitable to address both the (generic) BH solutions and the
Robinson-Bertotti ones 
(with an  $AdS_2\times S^2$ near horizon geometry) 
  reads
%
\begin{eqnarray}
\label{metric-generic}
ds= -a(r)^2  dt^2+ c(r)^2 dr^2+b(r)^2 d\Omega^2 , \qquad
\phi \equiv \phi(r), \qquad A= V(r) dt~,
\end{eqnarray}
with 
$(r,t)$  the radial and time coordinates, respectively, while
$d\Omega^2=d\theta^2+\sin^2\theta d\varphi^2$
is the usual metric on $S^2$.

The entropy of generic solutions (extremal or not) is
 fully accounted for by the Wald formula 
\cite{Wald:1993nt}
\begin{eqnarray}
\label{wald}
S&=&-2\pi A_H\,\frac{\delta {\cal L}}{\delta R_{\mu\nu\rho\sigma}}
\varepsilon_{\mu\nu}\varepsilon_{\rho\sigma}\nn
\\
&=&
2\pi A_H  \left [ -2 g^{\rho\mu}g^{\nu\sigma}
+2\alpha'a\,\e^{-\phi}\left\{-R^{\mu\nu\rho\sigma}
+4g^{\rho\mu}R^{\nu\sigma}+\left(-R+\Delta\phi\right)
g^{\rho\mu}g^{\nu\sigma}\right.\right.\nonumber
\\
&&
\left.\left.-g^{\rho\mu}\left(2\nabla^\nu\nabla^\sigma\phi
-\nabla^\nu\phi\nabla^\sigma\phi\right)
\right\}\right]
\varepsilon_{\mu\nu}\varepsilon_{\rho\sigma}\,,
\end{eqnarray}
where  $A_H$ is the event horizon area,  
$\varepsilon_{\mu\nu}$ is the binormal to the 
horizon of the BH, normalized so that 
$\varepsilon_{\mu\nu}\varepsilon^{\mu\nu}=-2$ and relation (\ref{wald})
is evaluated at the horizon.

\subsection{The $\alpha'=0$ limit: the GMGHS solution }
 
The GMGHS solution can be written in the form
(\ref{metric-generic}), with the metric functions
\begin{equation}  
\label{GHS-solution}  
a(r)^2=\frac{1}{c(r)^2}
=\left(1-\frac{r_+}{r} \right)\  \ , \qquad
b(r)^2=r^2 \left(1-\frac{r_-}{r}\right) \ , 
\end{equation}
and the Maxwell potential and dilaton field  
\begin{equation}
A=\frac{Q}{r}dt \ , \qquad 
e^{2\phi}=\frac{1}{2}\left(1-\frac{r_-}{r}\right) .
\end{equation}
The two free parameters $r_{+}$, $r_{-}$ 
(with $r_-<r_+$), corresponding to outer and inner horizon radius, respectively
They are related to the
ADM mass, $M$, and (total) electric charge, $Q$, by
\begin{eqnarray}   
  M = \frac{r_+}{2}  , \qquad 
	Q=\left(  
	\frac{r_-r_+}{2}
	\right)^{\frac{1}{2}} \ .
\end{eqnarray}
Other quantities of interest are the horizon area $A_H$ and
the Hawking temperature $T_H$ 
\begin{eqnarray} 
A_H=4\pi r_+^2\left(1-\frac{r_-}{r_+} \right) , \qquad 
T_H=\frac{1}{4\pi r_+}   \ .
\end{eqnarray}
The extremal limit, which corresponds to the coincidence limit $r_- = r_+$, 
results in a singular solution (as can be seen $e.g.$ by evaluating the Kretschmann scalar).
In this limit, the area of the event horizon goes to zero.
The Hawking temperature, however,  
approaches a constant. 

For completeness, let us mention  
 the existence of a magnetic version of the GMGHS solution,
which possesses the same metric, 
while the $U(1)$ potential and the dilaton are
\begin{eqnarray} 
A=Q\cos \theta d \varphi, \qquad e^{-2\phi}=\frac{1}{2}\left(1-\frac{r_-}{r}\right) ~.
\end{eqnarray}
However, this solution also becomes singular as $r_- \to r_+$.   A full duality orbit of solutions that interpolate between  these two  in Einstein-Maxwell-scalar models was recently discussed in~\cite{Herdeiro:2020iyi} 
(See also \cite{Shapere:1991ta, Ortin:1992ur}).

\section{Non-perturbative electrically charged solutions }

\subsection{The attractors}
\label{sec-at}
 
Taking into account  $\alpha'$ corrections, the possible existence  of non-perturbative generalizations of the extremal GMGHS
BHs 
with a nonzero horizon size
is suggested by the presence of a Robinson-Bertotti-type  exact solution, which we shall now discuss. 
The existence of such near-horizon geometry  is, moreover, closely  connected with the attractor mechanism~\cite{Sen:2004dp, Sen:2005kj, Sen:2005wa, 
Sen:2005iz}, as discussed above. We consider
  the following Ansatz, which is a particular case of (\ref{metric-generic}),
\begin{eqnarray}
   \label{att1}
ds^2=v_0^2 \left(\frac{dr^2}{r^2}-r^2 dt^2\right)+v_1^2 d\Omega_{(2)}^2, \qquad \phi(r)=\phi_0,\qquad V=q r,
\end{eqnarray}
where $v_0$, $v_1$, $\phi_0$ and $q$ are parameters
which satisfy a set of algebraic relations which 
result from the field equations.
The ansatz (\ref{att1}) was discussed by 
\cite{Sen:2005wa, Astefanesei:2006dd, Sen:2007qy} as 
the most general near horizon field configuration 
consistent with the SO(2, 1)$\times$SO(3) symmetry of 
AdS${}_2\times$S${}^2$.
Instead of following this route, however, 
in what follows we choose to determine the unknown  parameters by using the formalism proposed
in Refs. 
\cite{Sen:2005wa, Astefanesei:2006dd, Sen:2007qy},
 thus by extremizing an entropy function\footnote{   
{
We have verified that the same solution 
is recovered when solving the full set of covariant  field equations.
}
}.
 This alternative approach allows us to also  compute the entropy
of these BHs,
 and to show that the solutions exhibit attractor behaviour.

The entropy function is defined as
\begin{eqnarray}
\label{at2}
F(v_0,v_1,q,Q,\phi_0)=2 \pi \left( q Q-f(v_0,v_1,q, \phi_0) \right) \ ,
\end{eqnarray}
where 
$Q$ is the electric charge of the solutions,
while 
$f(v_0,v_1,q, \phi_0)$
is the Lagrangian density of the model (\ref{Ef})
evaluated for the Ansatz (\ref{att1})
and integrated over the angular coordinates,
\begin{eqnarray}
\label{at3}
f(v_0,v_1,q, \phi_0)
&=& \frac{1}{4\pi}
\int d\theta  d \varphi  \sqrt{-g} 
{\cal L}
\\
\nonumber
&=&
\frac{v_0^2-v_1^2}{2}
+e^{-2\phi_0}q^2 \frac{v_1^2}{2v_0^2}
-\frac{1}{4}\alpha' e^{-2\phi_0}
\left(
1-4 b\frac{e^{-4\phi_0}  q^4 v_1^2}{ v_0^6}
\right)~.
\end{eqnarray}
The attractor equations are:
\begin{eqnarray}
\label{at4}
  \frac{\partial F}{\partial v_0} & = & 0\,\,\,\Rightarrow~~~
 1=\frac{e^{-2\phi_0}  q^2 v_1^2}{ v_0^4}
 \left(
1+\frac{6\alpha' b e^{-4\phi_0}  q^2  }{v_0^4}
\right),
\\
\label{at5}
\frac{\partial F}{\partial v_1} & = & 0\,\,\,\Rightarrow~~~
1=\frac{e^{-2\phi_0}  q^2  }{ v_0^2} 
\left(
1+\frac{2\alpha' b e^{-4\phi_0} q^2  }{v_0^4} 
\right),
\\
\label{at6}
 \frac{\partial F}{\partial \phi_0} & = & 0\,\,\,\Rightarrow~~~
q^2=\frac{ \alpha'   v_0^2  }{2v_1^2}
\left(
1-\frac{12 b e^{-4\phi_0}   q^4 v_1^2 }{ v_0^6}
\right),
\end{eqnarray}
and
\begin{eqnarray}
\label{at7}
 \frac{\partial F}{\partial q} & = & 0\,\,\,\Rightarrow~~~
Q=\frac{  e^{-2\phi_0}  q v_1^2  }{v_0^2}
\left(
1+\frac{4 \alpha' b e^{-4\phi_0}  q^2  }{v_0^4} 
\right)
,
\end{eqnarray}
 The unique solution of the eqs. (\ref{at4})-(\ref{at7})
reads
\begin{eqnarray}
\label{at8}
v_0=  \frac{\sqrt{\alpha'} e^{-\phi_0}}{\sqrt{2}},~~~
v_1=\sqrt{\alpha'} e^{-\phi_0} \frac{\sqrt{2b}}{\sqrt{1+12b-\sqrt{1+16 b}}},~~~
q= \sqrt{ \alpha'}  \frac{\sqrt{\sqrt{1+16 b}}-1}{4\sqrt{b}}~,
\end{eqnarray} 
and
\begin{eqnarray}
\label{Q-expr}
Q= \sqrt{\alpha'} e^{-2\phi_0} \frac{\sqrt{b(1+16 b)(\sqrt{1+16 b}-1)}}
                               {1+12b-\sqrt{1+16 b}}~.
\end{eqnarray} 

Replacing the above expression and  back in the entropy function we obtain the remarkable simple expression of the entropy of an
extremal BH:
\begin{eqnarray}
\label{at11}
S_{\rm extremal}=  \alpha' ~e^{-2\phi_0}  \frac{2\pi \sqrt{1+16 b}}{3\sqrt{1+16 b}-1} ~,
\end{eqnarray}
which can also be expressed in terms of the electric charge,
\begin{eqnarray}
\label{at12}
S_{\rm extremal}= Q \sqrt{\alpha'}s_0\, \qquad {\rm with} \qquad s_0=\frac{\pi \sqrt{\sqrt{1+16 b}-1}}{2\sqrt{b}}  \ .
\end{eqnarray} 
As a check, we note that the result \label{ref} agrees with Wald’s form (3.11) evaluated
for the near horizon geometry.
Let us also remark
  that the constant $c$ which enters  (\ref{Ef}) does not enter the above expressions\footnote{
This is a consequence of the fact	the corresponding term in the Lagrangian density vanishes for 
the ansatz (\ref{att1}). }.
Moreover, one can easily see that (\ref{at8})
solves indeed the field equations, being a consistent solution.

As a final remark, we mention that there is no magnetic 
counterpart of the above solution. 
That is,
when employing  the same metric ansatz as in (\ref{att1}), a constant scalar and
 a magnetic $U(1)$ form 
$A=Q \cos\theta d\varphi$,
the field equations imply the following relation
\begin{eqnarray}
4 a \alpha'+v_1^2=0\ ,
\end{eqnarray}
which does not possess any physical solution.

\subsection{The bulk extremal BHs}

On general grounds, we expect that the above attractor solution describes the
neighbourhood of the event horizon of a bulk extremal BH.
In what follows, we give numerical evidence for the existence
of these configurations. 

\medskip

In the numerical study of the solutions,
it is convenient to
work in Schwarzschild-like coordinates, 
with the following choice in (\ref{metric-generic}):
$b(r)=r$,  $a(r)^2=e^{-2\delta(r)}N(r)$ and $c(r)^2=1/N(r)$, which results in the line element
\begin{equation}
\label{metric}
 ds^2=-e^{-2\delta(r)}N(r) dt^2+\frac{dr^2}{N(r)}+r^2 (d\theta^2+\sin^2\theta d\varphi^2)\ .
\end{equation} 

To construct RN-like extremal BH solutions, we assume the existence of a horizon located
at $r=r_H>0$.
In its exterior  neighbourhood, one finds the following 
approximate solution (which holds for an extremal BH only):
\begin{eqnarray}
\label{horizon1}
&&
N(r)=N_2(r-r_H)^2+\dots\ , \qquad
\delta(r)=\delta_0 + \delta_1 (r-r_H)+\dots\ ,
\\
&&
\nonumber
\phi(r)=\phi_0 + \phi_1 (r-r_H)+\dots \ , \qquad 
V(r)=v_1 (r-r_H)+\dots \ ,
\end{eqnarray}
with  
\begin{eqnarray}
\label{horizon2}
N_2=\frac{2}{\alpha'}\ , \qquad 
r_H= \frac{ e^{-\phi_0} \sqrt{\alpha'} \sqrt{2b}}{\sqrt{1+12b-\sqrt{1+16 b}}}>0 \ , \qquad 
v_1=\frac{e^{-\delta_0} \sqrt{ \sqrt{1+16 b}-1}}{2\sqrt{\alpha'}\sqrt{b}} \ ,
\end{eqnarray}
as imposed by the field equations.
Thus it turns out that only
the parameters 
$\phi_0$ 
and 
$\delta_0$
in the above near-horizon expansion
are essential,
the coefficients  $\phi_1$ and $\delta_1$ 
 being determined in terms of these.
However, 
their expression is too complicated to include here.

For large $r$, one finds the following asymptotic form of the solution:
\begin{equation}
\label{inf1}
N(r)=1-\frac{Q^2+ Q_s^2}{r}+\dots\ , 
\qquad \phi(r)=\frac{Q_s}{r}+\dots\ , \qquad 
V(r)=V_0+\frac{Q}{r}+\dots, \qquad 
\delta(r)=\frac{Q_s^2}{2r^2}+\dots\ .
\end{equation} 
The essential parameters in the above expansion
are the  mass $M$,
electric charges $Q$, electrostatic potential at infinity $V_0$
and scalar 'charge' $Q_s$.

These extremal BHs have finite global charges $M,Q$ as well as a finite
scalar 'charge' $Q_s$.
Their Hawking temperature vanishes, while the  entropy
takes a very simple form, as resulting from (\ref{wald})
 \begin{eqnarray}
\label{Ss}
S=S_E+S_{GB}\ ,\qquad  {\rm with} \qquad 
S_E=\frac{1}{4}A_H\ , \qquad  S_{GB}=\frac{1}{2} \alpha'  e^{-\phi_0}  \int_{H} d^2 x \sqrt{h} {\rm  R}~,
\end{eqnarray}
with 
${\rm  R}$ the Ricci scalar of the induced horizon metric:
\[
d\sigma^2=h_{ab}dx^a dx^b=r_H^2 (d\theta^2+\sin^2 \theta d\varphi^2) \ . 
\]
After replacing the near horizon expression of the solution ($e.g.$ $A_H=4\pi r_H^2$, ${\rm  R}=2/r_H^2$), 
the relation (\ref{at11}) is recovered\footnote{The expression of entropy becomes more complicated for non-extremal configurations,
with an extra-contribution to (\ref{Ss}).}.

We also note that
the equations of the model are invariant under the transformation
\begin{eqnarray}
\label{scale}
r\to \lambda r,~~~\alpha'  \to \lambda ^2 \alpha', 
\end{eqnarray} 
(with $\lambda>0$ an arbitrary  positive constant),
various  quantities of interest scaling accordingly, $e.g.$
\begin{eqnarray}
\label{scale2}
M\to \lambda M,~~~(Q ,Q_s)  \to \lambda ~(Q ,Q_s)  ~,
\end{eqnarray} 
such that only quantities invariant under (\ref{scale})
(like $Q/M$)
have a physical meaning.
In what follow, we use this symmetry to study solutions with
$\alpha'=0.5$
only,
without any loss of generality.

The metric functions  $N(r)$, $\delta(r)$, 
the scalar field $\phi(r)$
and the electric potential $V(r)$
are found by solving a set of four ordinary differential equations.
However, these equations are too long to display here, with hundreds of independent terms.
However, we mention that we have been able to find a suitable combination
of the field equations such that the functions $N,\delta$ still solve  first order equations,
while the functions $\phi$
and $V$
satisfy second order equations.
Moreover, the 2nd order
equation for the electric potential $V(r)$
possesses a 1st integral
which is used in practice to check the accuracy of the numerical results.

The solutions which smoothly connect the asymptotics
(\ref{horizon1}) and   (\ref{inf1})
are constructed numerically.
The only input parameters are
\begin{eqnarray} 
\{
 \alpha',~b,~c 
\},
\end{eqnarray}
while the constants $\phi_0,$ $\delta_0$
and 
$M$, $V_0$, $Q $,
$Q_s$
result from the numerical output,
with the electric charge $Q$ given by (\ref{Q-expr}),
a relation which provides an extra-test of the numerical accuracy.

\begin{figure}[t!]
\hbox to\linewidth{\hss%
	\resizebox{8cm}{6cm}{\includegraphics{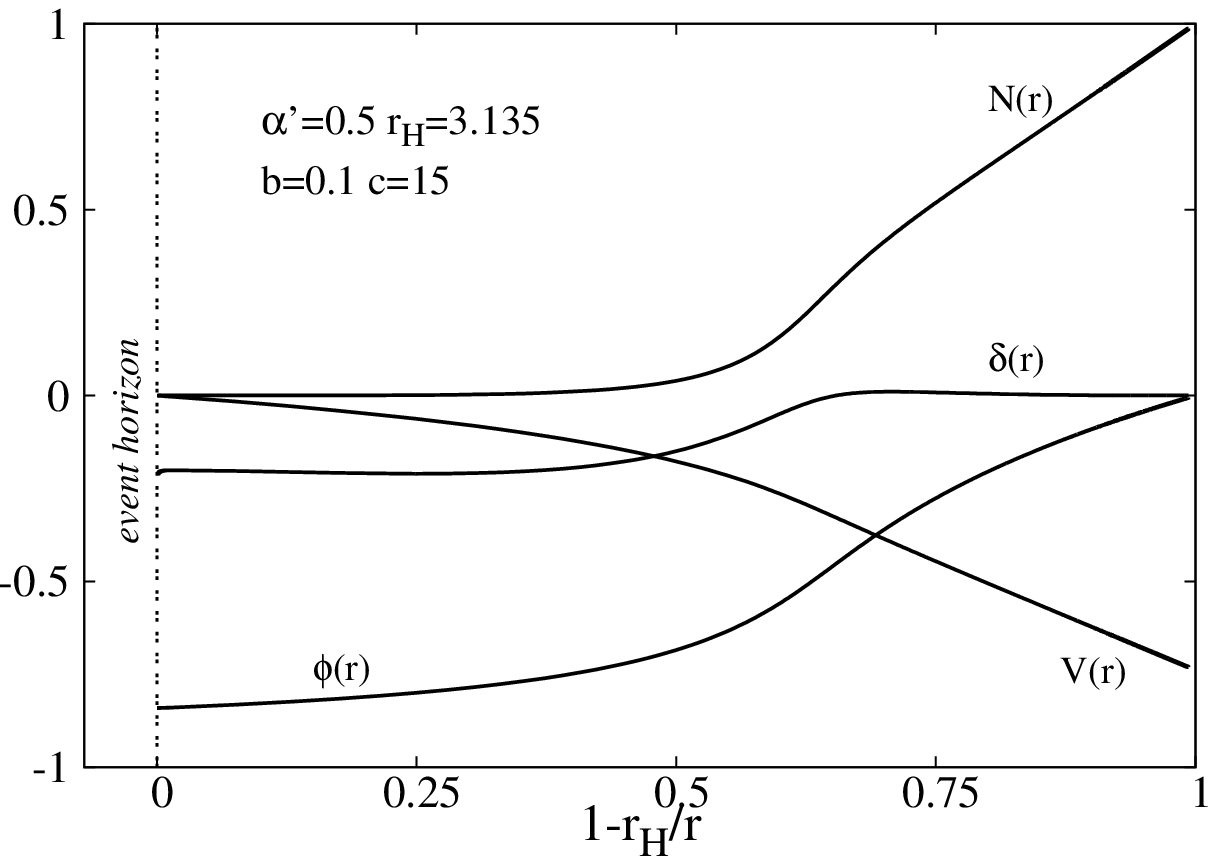}} 
		\resizebox{8cm}{6cm}{\includegraphics{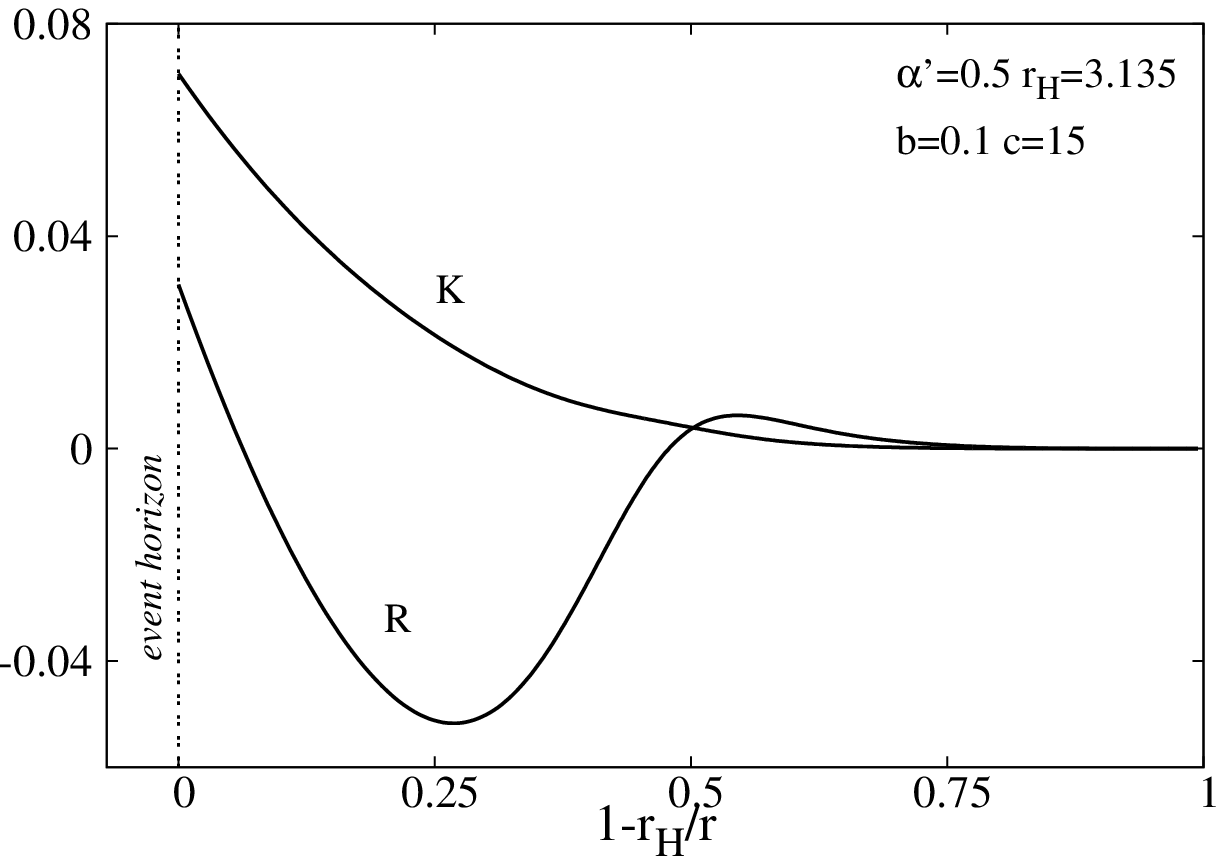}} 
\hss}
\caption{\small 
The profile functions (left) of an illustrative extremal BH solution 
are shown as functions of the compactified coordinate $1-r_H/r$.
The Ricci ($R$) and the Kretschmann ($K$) scalars are also shown (right). 
}
\label{profile}
\end{figure}

We follow the usual approach in such problems and, 
by using a standard ordinary differential equation solver, we evaluate the
initial conditions at  $r=r_H(1 +10^{-5})$ for global tolerance $ 10^{-14}$, adjusting for 'shooting' parameters 
$\phi_0$
and $\delta_0$,
and integrating towards $r\to \infty$,  looking for solutions with proper asymptotics.
The profile of a typical BH solution is shown 
in Figure \ref{profile}.
There, apart from the metric and matter functions,
we display also the Ricci and Kretschmann scalars,
which show that the geometry is regular on and outside the horizon.

 {\small \hspace*{3.cm}{\it  } }
%
 {\small \hspace*{3.cm}{\it  } }
\begin{figure}[h!]
\hbox to\linewidth{\hss%
\resizebox{8cm}{6cm}{\includegraphics{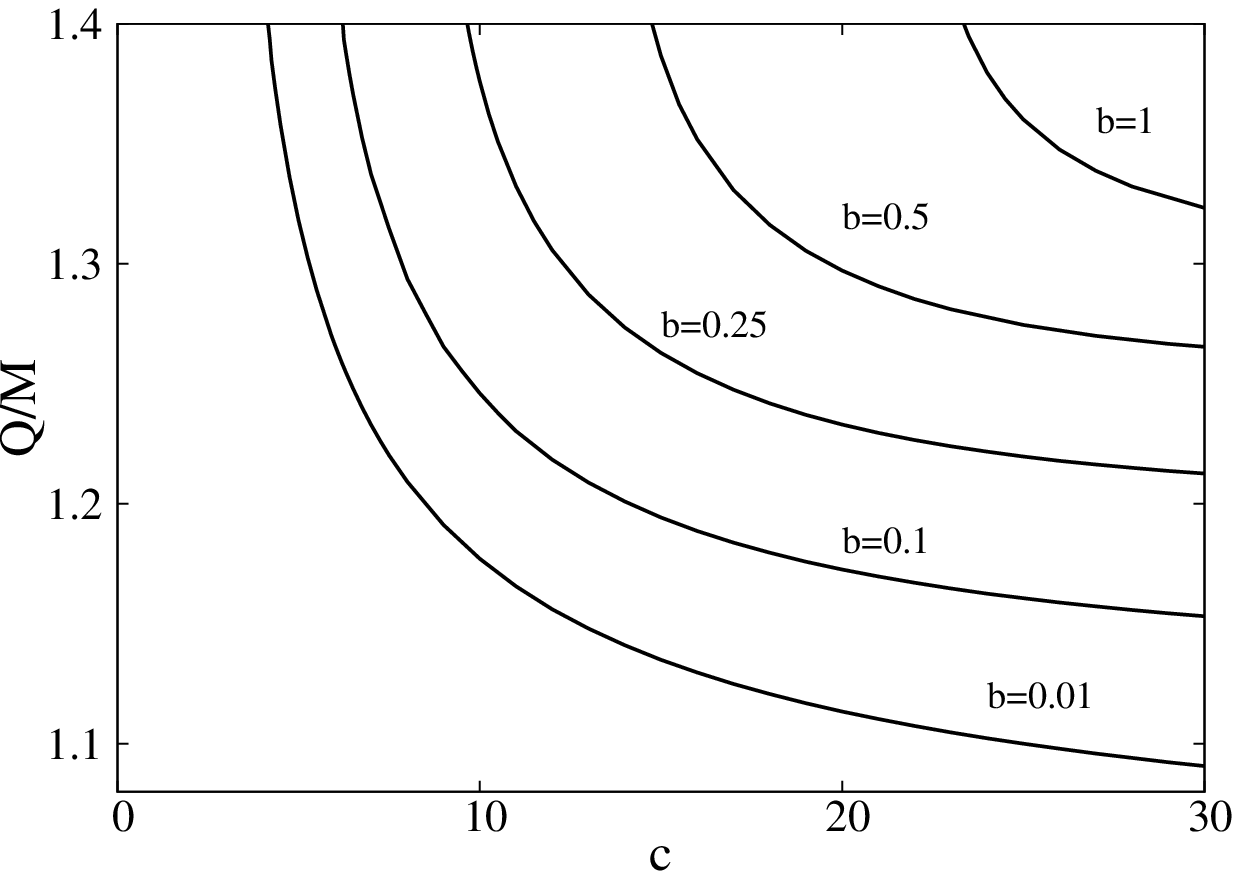}} 
  \resizebox{8cm}{6cm}{\includegraphics{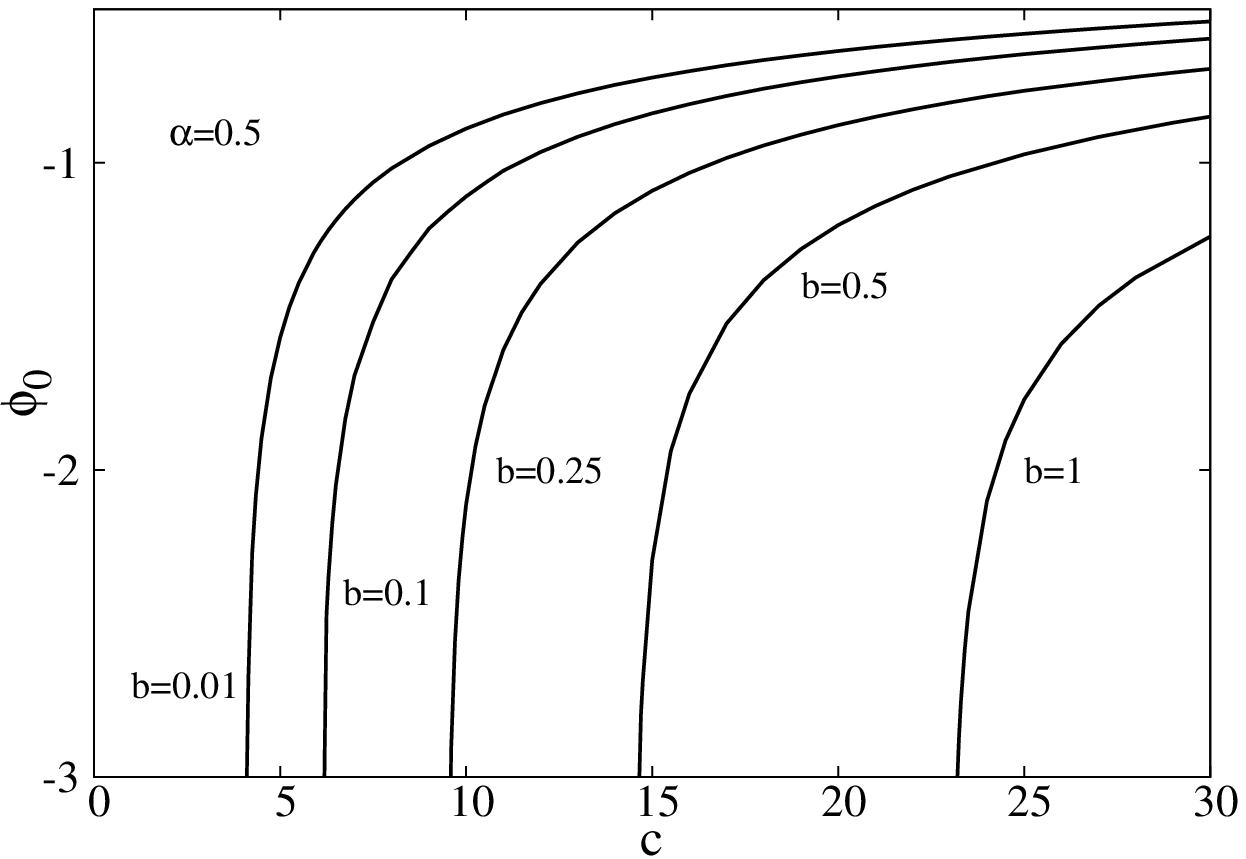}} 
\hss}
\caption{\small 
The charge to mass ratio (left) and the value of the dilaton field at the horizon (right) are shown as 
a function of the parameter $c$  for several values of $b$; $b,c$ are the two parameters which enter the (first order) $\alpha'$-corrections to the action.
}
\label{var-c}
\end{figure}

%
 {\small \hspace*{3.cm}{\it  } }
\begin{figure}[h!]
\hbox to\linewidth{\hss%
\resizebox{8cm}{6cm}{\includegraphics{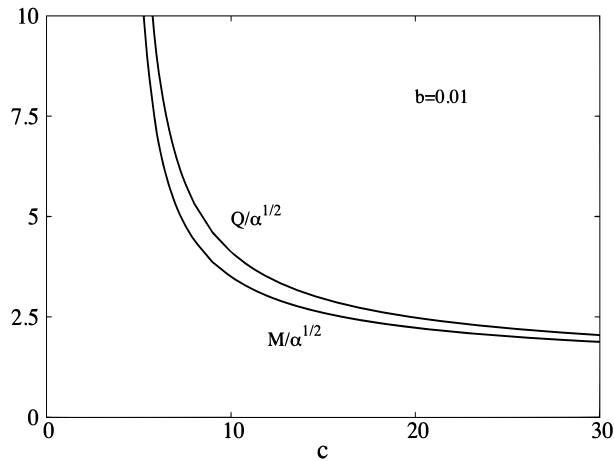}} 
\hss}
\caption{\small 
The BH mass $M$ and electric charge $Q$ 
are shown to decrease as 
the parameter $c$ increases, for $b=0.01$\,. 
If we compare it with the figure \ref{var-c}, 
we observe that the charge-to-mass ratio $Q/M$ 
decreases as $M$ decreases.
}
\label{extra}
\end{figure}

The basic properties of the extremal solutions can be summarized as follows. 
First, given the input constants $(\alpha',~b,~c)$,
regular BH solution appears to exist for a single
set 
of the 'shooting' parameters
$(\phi_0,\delta_0)$,
while the profile of the scalar field
$\phi(r)$
is nodeless.
Thus it is natural to conjecture that, similar to the 
$\alpha'= 0$
case in
\cite{Gibbons:1987ps,Garfinkle:1990qj}, 
only nodeless solutions exist also for the model in this work.

 Second, for a given $b$,
the existence of solutions depends on the value of 
input constant $c$, which enters the $\alpha'$-term in the action (\ref{Ef})
(we recall that $c$ is absent in the attractor solution).
Taking $\alpha'=1/2$,
we have considered various values of the constant $b$
and varied the value of the parameter $c$.
Then no upper bound on $c$ was found.
However, for a given $b$,
 the solutions stop to exist for $c<c_0$, with $c_0$
increasing as $b$ increases (see Figure \ref{var-c}, both  panels).
As $c\to c_0$,
the value of the scalar field at the horizon  appears
to increase (in modulus) without bound, and the solution becomes singular, 
 a feature which cannot be predicted by the attractor analysis in Section 3.1.
A detailed study of the critical behaviour in this limit may be of  interest, but  it is outside the scope of this work.

Given the above remark,
the existence of 
regular solutions with $c=2$ 
(as implied by the perturbative results in Ref.
\cite{Natsuume:1994hd})
seems  to require very small values of the parameter $b$, of the order $10^{-5}$ or smaller.
Unfortunately, the numerical accuracy deteriorates in that region of the parameter space and we could 
not explore this case.

Finally, perhaps the most interesting feature of the solutions found so far is that the ratio
$Q/M$ is always greater than one - see Figure \ref{var-c}, left panel.
But the charge-to-mass ratio $Q/M$ 
decreases when $c$ increases for fixed $b$. 
On the other hand, from Figure \ref{extra}, 
we find that as the parameter $c$ grows, 
the BH mass $M$ decreases for fixed value of $b$\,. 
This feature seems to be at tension with the rationale of the weak gravity conjecture.

\section{Discussion}
In  this  work we have confirmed that $\alpha'$  corrections can de-singularize the extremal GMGHS solution, an influential  stringy BH whose extremal limit is long known to be singular. This gives an illustrative example of how higher order corrections motivated by quantum gravity  can be key (and non-negligible) to understand the BH geometry  on  and outside a horizon. 
These higher-curvature terms also contribute to the global charges and energy of the system 
\cite{Cano:2018qev, Cano:2018brq, Faedo:2019xii}.

The BH solutions with $\alpha'$ corrections 
in string theory, moreover, shows the importance of corrections in  the Riemann tensor (rather than,  say, just in the Ricci  scalar  as in $f(R)$ models); taking them into account is 
fundamental because the curvature scalars do not capture 
all the possible terms in the equations of 
motion at higher orders in $\alpha'$ corrections 
\cite{Chimento:2018kop}. 
Here  we have focused on static BHs  but there are also studies with rotating BHs with first order 
correction in $\alpha'$ \cite{Campbell:1990ai, Campbell:1990fu, 
Mignemi:1992pm} - see also the related studies in \cite{Yunes:2009hc, Yunes:2011we, Pani:2011gy, Ayzenberg:2014aka,
Cano:2019ore}.

We observe that the charge-to-mass ratio $Q/M$ 
decreases when the BH mass $M$ decreases for fixed value 
of $b$\,, which is different from previous examples 
\cite{Cheung:2018cwt, Cheung:2019cwi}.
Whether the charge-to-mass ratio increases or decreases depends 
on the particular structure of the higher derivative terms.
If it increases then BHs can decay to smaller BHs,
while such decays are forbidden in the other case. Although we get
$Q/M>1$, the charge-to-mass ratio decreases as the mass decreases.
Hence, our results do not assure that an
extremal BH is always able to decay to smaller extremal BHs of marginally higher charge-to-mass ratio. 

 Let also remark that
 the results reported here are exploratory, but establish a proof of concept that such higher order corrections lead to BH solution that are non-singular on  the horizon  and can be extended throughout  the whole spacetime. 
We have also confirmed the existence of 
non-extremal BH solutions of the same model.
These BHs possess very similar 
far field and near horizon expressions,
 the main difference being that the function $N(r)$
possesses a single zero as $r\to r_H$,
\begin{eqnarray}
\label{horizonNE}
N(r)=N_1(r-r_H) +\dots ~,
\end{eqnarray}
(also the value $r_H$ is not fixed in this case). 
{ 
Moreover, 
we have found clear numerical evidence that  
the non-extremal BH solutions  exist for a large part of the $(b,c)$-plane.
We hope to return elsewhere with a detailed study of these configurations.
}

As a final intriguing  remark, it  would be interesting  to clarify the absence of  a magnetic counterpart to the electric attractor in Section~\ref{sec-at}, and  its implication in  the contex  electric/magnetic duality for the  $\alpha'$ corrected theory.

\bigskip

\noindent {\large\bf Acknowledgements}
\\  
The  work of E.R.  is  supported  by  the Center  for  Research  and  Development  in  Mathematics  and  Applications  (CIDMA)  
through  the Portuguese Foundation for Science and Technology (FCT - Fundacao para a Ci\^encia e a Tecnologia), 
references UIDB/04106/2020 and UIDP/04106/2020, and by national funds (OE), through FCT, I.P., 
in the scope of the framework contract foreseen in the numbers 4, 5 and 6 of the article 23,of the Decree-Law 57/2016, of August 29, changed by Law 57/2017, of July 19.  
We acknowledge support  from  the  projects  PTDC/FIS-OUT/28407/2017,  CERN/FIS-PAR/0027/2019 and PTDC/FIS-AST/3041/2020.  
 This work has further been supported by the European Union’s Horizon 2020 research and innovation (RISE) programme H2020-MSCA-RISE-2017 Grant No. FunFiCO-777740.  
The authors would like to acknowledge networking support by the COST Action CA16104.
The work of K. U. is supported by Grants-in-Aid from the Scientific 
Research Fund of the Japan Society for the Promotion of 
Science, under Contract No. 16K05364.

\section*{Appendix}
\appendix
In this appendix, we work out the explicit form of the 
components of the Einstein equations for a metric 
(\ref{metric-generic}) which is compatible with the expected 
form of the sought for attractor solution.  
We refer to section 3 for the physical motivation leading 
to a metric of the form. Using again the expressions 
(\ref{metric}), 
Einstein equations become
\begin{eqnarray}
\label{eqN}
&&
r N'+N-1+r^2N\phi'^2+e^{2(\delta-\phi)}r^2 V'^2
+\alpha' e^{-2\phi} N 
                                  \bigg( 
4a P_{N}
\\
\nonumber
&&
{~~~~~~~~~~~~~~~~~~~~~~~~~~~~~~~~~~~~~~~~~~}
+6b e^{4(\delta-\phi)}\frac{r^2 V'^4}{N}
-c e^{2(\delta-\phi)}r^2 \phi'^2 V'^2
                                  \bigg)
																				=0,
																				\\
\label{eqd}
&&
\delta'+r \phi'^2+\alpha' e^{-2\phi }\left(4 P_{\delta}a+c e^{2(\delta-\phi)} r\phi'^2 V'^2 \right) =0,
 \\ 
%
\label{constr}
&&
\delta''-\frac{N''}{2N}+\delta'
\left(\frac{1}{r}-\delta'\right)+\frac{N'}{N}
\left(
\frac{3\delta'}{2}-\frac{1}{r}
\right) -\phi'^2+\frac{e^{2(\delta-\phi)} V'^2}{N}
\\
\nonumber
&&
{~~~~~~~~~~~~~~~~~~~~~~~~}
+\alpha' e^{-2\phi} 
\left(
\frac{4a}{r}P_c+\frac{2b e^{4(\delta-\phi)}V'^4}{N}
-c e^{2(\delta-\phi)}\phi'^2 V'^2
\right)=0\,,
\end{eqnarray}
%
%
while the equations of motion for scalar and gauge fields are
\begin{eqnarray} 
&&
\label{eqPhi}
\phi''+\left(\frac{2}{r}+\frac{N'}{N}-\delta'\right)\phi'
-\frac{e^{2(\delta-\phi)} V'^2}{N}
+\alpha' e^{-2\phi}
\bigg(
2 a P_{\phi}-b\frac{6 e^{4(\delta-\phi)}V'^4 }{N}
\\
\nonumber
&&
{~~~~~~~~~~~~~~~~~~~~~~~}
+c e^{2(\delta-\phi)} V'
(
(\frac{2}{r}+\delta'+\frac{N'}{N}-2\phi')\phi' V'
+V'\phi''+2\phi'V'
)
\bigg) =0,
\\
\label{eqV}
&&\left
(e^{\delta-2\phi}r^2 V'
(1+\alpha' e^{-2\phi}(4 b e^{2(\delta-\phi)}V'^2-c N \phi'^2
)
)
\right
)'=0,
\end{eqnarray}
%
where a prime "$'$" states a derivative with respect to 
the radial coordinate $r$\,. 
The equation (\ref{eqV}) for the electric potential 
$V(r)$ consists of a first integral. 
Also, in order to simplify the relations, we have defined 
the following quantities
\begin{eqnarray}
\nonumber
&&
 P_N=N' \phi'\left( 3(1+r^2 \phi'^2)-\frac{1}{N} \right)
+2\phi'^2
\bigg(
3(1+rN')
\\
\nonumber
&&
{~~~~~~~~~~~~~~}
-N(1+r\phi'(4+3 r\phi')) 
\bigg)
+2\phi''(N-1+r N \phi'(4+3 r\phi')),
\\
\nonumber
&&
P_\delta=\left(
\frac{3N-1}{r}+3N\phi' (2+r\phi')
\right)\delta'\phi'
+2 \phi'^2 \bigg(
\frac{2(1-N)}{r}-N\phi'(5+3r \phi')
\bigg)
\\
\nonumber
&&
{~~~~~~~~~~~~~~~~~~}
+ \phi'' \left(
\frac{N-1}{r}+N\phi'(4+3r \phi)
\right),
\\
\nonumber
&&
P_\phi= \frac{1}{r^2}
\bigg(
\left(2(1-N)\delta'+5N'-4(1-3N)\phi' \right)\delta'
-12 N'\phi'-
\left( 3\delta'+N'-4\phi'  \right)\frac{N'}{N}
\bigg)
\\
\nonumber
&&
-\frac{4}{r}
\bigg(
\delta'(2N\delta'-5N')+\frac{N'^2}{N}
\bigg)\phi'
+
\bigg(
-\frac{4(1+2N)}{r^2}+ (2N(\frac{8}{r}-3\delta')+15 N')\delta'
-(\frac{26}{r}+\frac{3N'}{N})N'
\\
\nonumber
&&
+12 \phi'(N\phi'-N')
\bigg)\phi'^2
+2
\bigg(
\frac{N-1}{r^2}+\frac{4N \phi'}{r}+3N \phi'^2
\bigg)\delta''
+
\bigg(
\frac{1-N}{r^2N}-\frac{4\phi'}{r}-3\phi'^2
\bigg)N''
\\
\nonumber
&&
+\frac{2}{r}
\left(
\frac{2(1-N)}{r}+4N\delta'-2N'-3(rN'+2N(2-r\delta'+2r \phi'))\phi')
\right) \phi'',
\\
\nonumber
&&
P_c=(5N'-2N\delta')\delta'\phi'-
\left(N'(2-5r \delta'+r\phi')+2N (r\delta'^2+\delta'(1+2r\phi')
-\phi'(2+3r\phi')) \right)\phi'^2
\\
\nonumber
&&
+2N\phi' (1+r \phi')\delta''-\phi'(1+r \phi')N''
+\bigg(
-N'(1+2r \phi')+2N (\delta'+2(-1+r\delta')\phi'
\\
\nonumber
&&
{~~~~~~~~~~~~~~~~~~~~~~}
-3 r\phi'^2)
\bigg)\phi'' -\frac{N'^2\phi'(1+r \phi')}{N}.
\end{eqnarray}
The component of Einstein equations displayed in (\ref{constr}) 
is the constraint which will be used when searching for 
black hole solutions, including the extremal configuration.
Here the functions $N, \delta$ solve first order equations,
while the functions $\phi$ and $V$ 
satisfy second order equations, which was used in our
 numerical treatment of the problem. 
%


	
\end{document}